\documentclass{article}
\usepackage{graphicx}
\graphicspath{ {./Figures/} }
\usepackage{siunitx}
\usepackage{authblk}
\usepackage{soulpos}
\usepackage{wasysym}
 
\title{Femtosecond laser written waveguides in sapphire for visible light delivery}
\author[1,2]{Sarah Winkler}
\author[2]{Joachim R. Krenn}
\author[1,3]{Jakob Wahl}
\author[1,2]{Alexander Zesar}
\author[1]{Yves Colombe}
\author[1]{Klemens Schüppert}
\author[1]{Clemens Rössler}
\author[4]{Christian Sommer}
\author[4]{Philipp Hurdax}
\author[4]{Philip Lichtenegger}
\author[4]{Bernhard Lamprecht}
\affil[1]{Infineon Technologies Austria AG, Siemensstraße 2, A-9500 Villach, Austria}
\affil[2]{Institut für Physik, Universität Graz, 8010 Graz, Austria}
\affil[3]{Institut für Experimentalphysik, Universität Innsbruck, 6020 Innsbruck, Austria}
\affil[4]{MATERIALS - Institute for Sensors, Photonics and Manufacturing Technologies, JOANNEUM RESEARCH, Franz-Pichler-Straße 30, 8160 Weiz, Austria}
\date{July 2023}

\begin{document}

\maketitle

\noindent

\section{Abstract}
A promising solution for scalable integrated optics of trapped-ion quantum processors
    are curved waveguides guiding visible light within sapphire bulk material.
To the best of our knowledge,
    no curved waveguides were investigated in sapphire so far,
    and no measurements of waveguides with visible light in undoped planar sapphire substrates were reported.
Here, we demonstrate femtosecond laser writing of depressed cladding waveguides in sapphire.
Laser parameters, such as
    pulse energy, pulse duration, and repetition rate, as well as waveguide geometry parameters,
    were optimized to guide 728 nm light.
This resulted in single-mode waveguides with a propagation loss of $1.9(3)$ dB/cm.
The investigation of curved waveguides
    showed a sharp increase in total loss for curvature radii below 15 mm.
Our results demonstrate the potential of femtosecond laser writing as a powerful technique
    for creating integrated optical waveguides in the volume of sapphire substrates.
Such waveguides could be a building block for integrated optics in trapped-ion quantum processors.

\section{Motivation}
In trapped ion quantum computing, simulation, communication, and sensing,
    ions are the elementary building blocks for quantum information processing and high-precision measurements \cite{Haeffner2008, Bruzewicz2019, Blatt2012, Brewer2019}.
Individual ions are confined by electromagnetic fields,
    and their internal (electronic) and motional states are manipulated with laser beams.
Typically, free-space optics are used to perform quantum operations such as
    state preparation, readout, single-ion gates, and multi-ion entangling gates.
However, this approach has significant drawbacks, such as beam instability, high cost, and limited scalability.

A promising alternative to overcome these drawbacks is the integration of waveguides into an ion trap \cite{Mehta2016,Day2021,McGuinness2022}.
In comparison to free-space optics,
    these approaches simplify the overall design of the quantum processing hardware 
    and reduces the number of components that require regular alignment,
        thereby allowing for a stable system \cite{Niffenegger2020,Vasquez2023}.
Therefore, the integration of photonic components can be crucial 
    to realize large-scale trapped ion quantum processors,
    where scalability and compactness are critical.

Femtosecond (fs) laser writing \cite{Miura1997, Osellame2012} is a powerful technique
    to produce integrated optical waveguides in the volume of transparent substrates.
In this process, a fs-laser beam is tightly focused in the substrate.
While the substrate material is transparent for the wavelength of the fs laser,
    the high peak intensity at the laser focus allows  photons to cross the bandgap via multi-photon  absorption or tunneling ionization. This generation of free charge carriers enables avalanche ionization, causing the energy from the beam to be rapidly transferred to the electrons in the conduction band. The resulting free electron plasma then transfers its energy to the crystal lattice,
    leading to localized structural modifications \cite{Osellame2012}.
Shifting the focal volume in 3D
    enables the fabrication of arbitrary structures with a high spatial resolution
    as set by the focal volume in the \SI{}{\micro\meter} range.

Fs-laser writing can be used to create waveguides of different geometries and dimensions,
    including straight and bent waveguides,
    as well as waveguides with refractive index profiles that can be tailored to specific applications \cite{Tan2021, Sun2022}.
The technique allows for the creation of complex optical circuits and devices \cite{Corrielli2021}.
The method is applicable to a wide range of transparent materials,
    including glass \cite{Dong2013} and dielectric crystals \cite{Ren2016, Berube2018, Romero2019, Li2022}.
Importantly, fs-laser writing is a non-contact and non-destructive, maskless and rapid-prototyping technique,
    which allows for the creation of optical structures with high precision and reproducibility.
Additionally, it can be integrated with other microfabrication techniques,
    such as FLICE (fs-laser irradiation and chemical etching) or
    fs-laser ablation to create complex optical structures on a single substrate \cite{Li2022}.

In general, for fs-laser-modified volumes,
    three types of morphological changes are observed depending on laser parameters and material properties \cite{Osellame2012}.
In the regime of low absorbed energy, smooth changes in the refractive index are created.
The intermediate, medium absorbed energy regime leads to birefringent refractive index modifications.
High energy modifications include the formation of empty voids due to micro-explosions.
Neither medium nor high-energy modifications are desirable for optical guiding. 

\begin{figure}[t]
\includegraphics[width=1\textwidth]{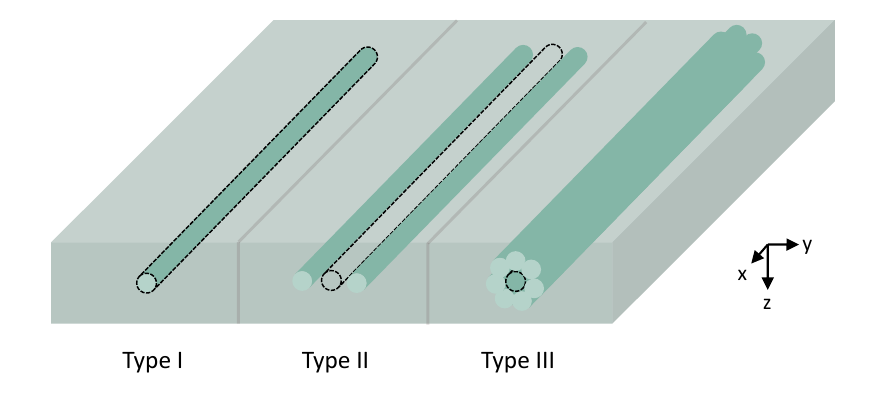}
\caption{Types of fs-laser-written waveguide geometries. Type I is made of a single fs-laser-written line (green area) which is congruent to the waveguide core (dotted area). Type II: Two parallel fs-laser-written lines enclose the waveguide core in between them. Type III consists of multiple lines surrounding the waveguide core. Adapted from \cite{Chen2013}.}
\centering
\label{fig:WG-types}
\end{figure}

Depending on the relative position of the waveguide core with respect to the fs-laser-written lines,
    three different types of waveguide geometries are defined \cite{Corrielli2021}.
Type I geometries are based on a positive change in refractive index,
    allowing direct writing of waveguides as the irradiated focal volume acts as the waveguide core (Fig. \ref{fig:WG-types} left).
This type is common in glasses,
    but rarely met in crystals where the refractive index change is usually negative \cite{Li2022}.
Type II geometries are characterized by positive refractive index changes,
    which are located in the vicinity of the laser-written lines (Fig. \ref{fig:WG-types} middle).
For type II geometries, the track has a lower index and cannot be used as a waveguide;
    however, the expansion of volume in the focal plane gives rise to a compression of its vicinity, 
    leading to the refractive index increments in the surrounding of the track \cite{Li2022}.
Usually, these waveguides are written in a dual-line geometry,
    where the waveguide core is located in between two laser-written lines.
Type III geometries are defined by negative refractive index changes as well,
    but significantly differ from the type II dual-line geometry.
In type III geometries,
    the unexposed waveguide core is surrounded by many low index lines close to each other acting as a low-index barrier
        that confines the guided light inside (Fig. \ref{fig:WG-types} right).
This waveguide geometry is also referred to as depressed cladding waveguides (DCWs) 
    \cite{Berube2018,Dong2013,Romero2019}   
    and this type of waveguide geometry is explored in this article. 

Sapphire is a material of choice for macroscopic and microfabricated ion traps \cite{Hempel2014, Hellwig2010, Daniilidis2011, Allcock2013, Kunert2013}
    due to its low radio-frequency loss tangent and its high thermal conductivity compared to most dielectrics.
These properties ensure that the trap temperature increases only minimally
    while applying radio-frequency voltages to the trap electrodes,
    which is desirable in particular for quantum computing and sensing applications.
Since sapphire is transparent to visible light
    due to its high bandgap of 8.7 eV,
    it is also a promising material for microfabricated, optical waveguides integrated into ion traps.
Fs-laser modification of sapphire results in morphological changes in the crystal structure of the material. The intense light pulses cause a nonlinear interaction with the crystal lattice,
    leading to a localized modification of the sapphire \cite{Ren2016}.
    These modifications typically lead to a decrease of the refractive index
    \cite{Li2022}
    which is the basis of the fabrication of a waveguide with a lower refractive index cladding surrounding the unmodified sapphire
material. 
For one such waveguide inscribed using 515 nm light in sapphire, the magnitude of the refractive index change has been found to be $-2.5\cdot 10^{-3}$ \cite{Berube2019}.

Numerous articles on waveguides written in Ti-doped sapphire have been published \cite{Bai2012},
    as doped sapphire is a highly attractive laser material.
In 2018, Bérubé et al. \cite{Berube2018} reported the first waveguides in undoped sapphire substrates.
Their waveguides were based on a depressed cladding structure
    and showed single mode behavior for 2850 nm and a propagation loss of 0.37 dB/cm.
In April 2022, Wang et al. \cite{Wang2022} reported the fabrication of a single-mode fiber Bragg grating in sapphire.
In a preceding report,
    they described the successful fabrication of DCW in undoped sapphire
    and  a total loss for a 1-cm long DCW of 6.09 - 6.68 dB (at 1550 nm),
    depending on the polarization direction of the in-coupling light.
Also in April 2022,
    Kefer et al. \cite{Kefer2022} reported the fabrication of single mode waveguides in arbitrarily oriented, undoped sapphire crystals 
    based on a photonic crystal approach achieving 1 dB/cm loss at 1550 nm.
Thus, to the best of our knowledge,
    no curved waveguides were investigated in sapphire so far
    and no measurements with visible light of waveguides in undoped planar sapphire substrates were reported.

\section{Methods}
\subsection{Femtosecond laser writing}
\begin{figure}[t]
\centering
\includegraphics[width=0.8\textwidth]{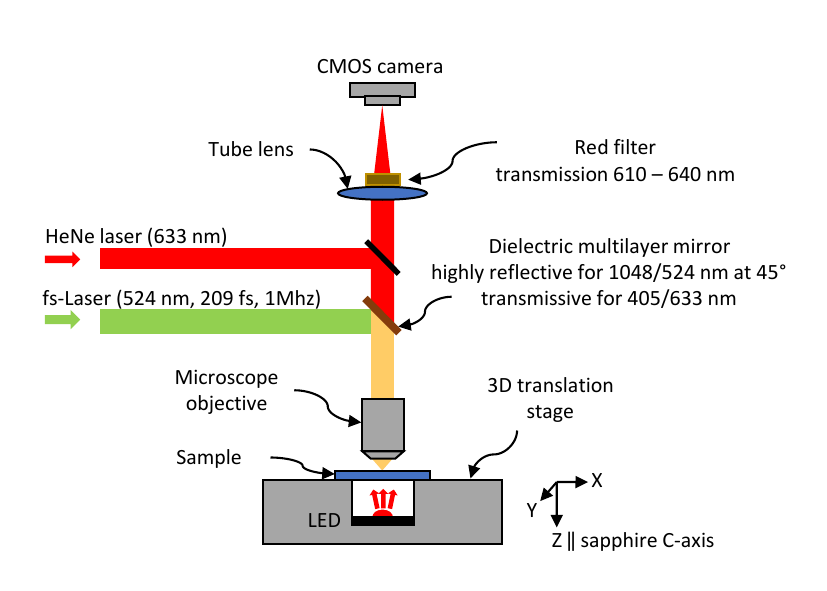}
\caption{Optical setup of the laser lithography platform. A fs-laser is widened and collimated before it enters the setup. Here it is superimposed with a HeNe laser, which is visible in the imaging setup and used for z-levelling. The sapphire crystal C-axis is parallel to the z-axis in this setup.}
\centering
\label{fig:Writing-setup}
\end{figure}

We write the optical structures in C-cut sapphire
    using a Gaussian laser beam focused inside the bulk material.
Moving the sample in 3D relative to the laser focus
    allows for precise creation of complex structures within the sapphire.
The writing direction is perpendicular to both the laser beam propagation direction and the C-axis of the sapphire crystal,
    and the cross-section dimensions of the tracks of modified refractive index are determined by the diffraction-limited waist radius and the depth of focus of the laser beam, respectively.
Fig. \ref{fig:Writing-setup} shows a schematic drawing of the laser lithography system.
The setup allows two laser beams to run collinearly with each other: a fs-laser beam for the multi-photon structuring process and a red HeNe laser for alignment.
The laser platform allows the use of a wide range of focusing optics with numerical apertures ranging from 0.25 to 1.4.
Femtosecond laser (fs-laser) structuring is performed with the second harmonic (524 nm) of a Spectra-Physics Spirit fs-laser. The second harmonic is used, because shorter wavelengths (524 nm instead of 1048 nm) 
allow for tighter focusing due to diffraction limits being inversely proportional to wavelength. 
As a result, using a green laser could achieve higher spatial resolution in the inscription process, 
allowing for finer waveguides with smaller mode field diameters, which have the advantages of (1) lower cutoff wavelengths, i.e. being able to guide shorter wavelength light, which is of interest for ion manipulation and (2) allowing for smaller curvature radii due to the stronger confinement associated with a higher angle of incidence at the core-cladding interface.
The pulse duration of the fs-laser
can be set between 209 fs and 9 ps and a pulse repetition rate of maximum 1 MHz 
        that can be divided down to single pulses.
The fs-laser beam path includes an attenuator for external power control (half-wave plate and polarizing beam splitter),
    which is necessary to adjust the required maximum average power.
Furthermore, the average output power of the laser is adjusted via an external 0-5 V signal, 
    allowing for rapid power adjustments during laser writing.
A beam expander before the optical setup shown in Fig. \ref{fig:Writing-setup} ensures full and uniform illumination of the focusing objective.
The polarization is adjustable from circular polarization to any linear polarization.

Structures are written into the sapphire by moving the substrate in the horizontal xy-plane,
    which is perpendicular to the axis of the focusing optics (z-direction).
The xy translation is achieved using a pair of linear motion stages (Newport XML210-S),
with a minimum incremental movement of 1 nm, a typical bidirectional repeatability of \SI{\pm32}{\nano\meter} and a travel range of \SI{210}{\milli\meter}.
The writing setup, depicted in in Fig. \ref{fig:Writing-setup}, is placed on a granite base to minimize the effects of external vibrations.
The optical assembly, comprising focusing optics and opto-mechanical components, is mounted on a breadboard
    that can be moved using a z-stage to adjust the focus of the writing beam.
The z-stage is a high-load vertical linear stage from Newport (IMS100V) 
    with a typical bidirectional repeatability of \SI{\pm150}{\nano\meter} 
    and a vertical travel range of \SI{100}{\milli\meter} for loads up to \SI{400}{\newton}.
Fine adjustment is achieved by an additional piezo z-stage on which the microscope objective is mounted.
The linear stages are controlled by a Newport XPS-D motion controller. 

An integrated inline video microscope with reflected and transmitted light illumination by means of a HeNe laser (633 nm) 
    and two LEDs is used for focusing, leveling, alignment and real-time process monitoring. The HeNe-laser beam is used for focusing by imaging the reflected light from the optical interfaces of the sample onto the integrated camera.
Correct adjustment of the focus of the writing beam is an important prerequisite for precise laser patterning.
In addition, this method also makes it possible
    to align or level the sample horizontally via a tilt stage.
This is necessary 
    to remain parallel to the substrate surface 
        when scanning over larger areas,
    meaning that the best possible focusing conditions are maintained even
        when scanning over centimeters.
For process monitoring,
    the substrate is illuminated by a red LED underneath the transparent substrate.
Since the laser exposure leads to a change in the optical properties of the structured areas (refractive index modification),
    the effect of laser structuring becomes visible in transmission light microscopy.
The setup enables real-time monitoring of the laser patterning process and its quality.
To prevent saturation of the camera with ambient light, filters are placed in front of the camera, which only transmit red light in the range of the LED and HeNe emission of approx. 610 to \SI{640}{\nano\meter}. 

An essential point of the experiment concerns the creation of so-called writing trajectories for the waveguide writing.
We use a specific operating mode of the Newport XPS motion controller, known as PVT (Point-Velocity-Time) trajectories.
In PVT, each individual path element is defined by the relative displacement (P), the final speed (V) and the duration for the path element (T) for all motion axes in the common motion group.
An additional pass-through card for the Newport XPS motion controller (XPS-DRV00P) serves as a source for analog voltage signals.
The card is configured together with the other motion axes (x-, y-, z-, z-Piezo) in a multi-axis motion group,
    providing synchronized signals aligned with the positions of the motion axes.
The individual path elements for all involved motion axes are defined by the user in a text file,
    which is sent to the XPS controller.
The controller then calculates and executes the trajectory by interpolating a smooth cubic function
    that traverses all specified positions at the defined times and velocities.
Since a software for creating the PVT trajectory files is not commercially available,
    the necessary PVT files are generated by self-written Python scripts,
    implemented in Rhinoceros 3D (McNeel Europe), a 3D modeling software.
Together with other important structuring parameters,
    such as laser writing velocity and the analog input for pulse energy,
    a text file is created
    which defines the PVT trajectory,
    required by the motion controller for movement in x-, y- and z-directions and for the laser power controlled by the analog signal.

\subsection{Laser-writing of depressed cladding waveguides}
\begin{figure}[t]
\includegraphics[width=1\textwidth]{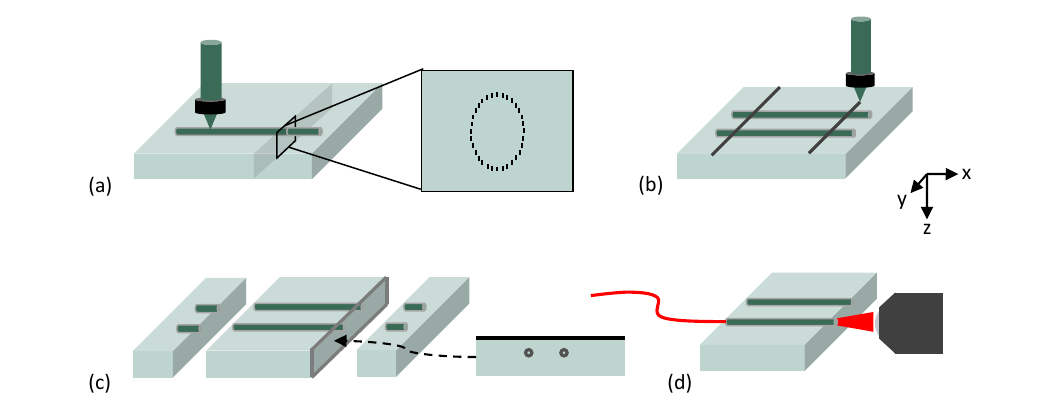}
\caption{Schematic representation of 3D laser-writing of depressed cladding waveguides.
    (a) Constructing the DCW cladding with single scans arranged in a circle or ellipse (black dots in inset). Process repeated for multiple lines with different parameters,
    (b) laser ablation, 
    (c) exposing end facets for further characterization via manual cleaving,
    (d) measurement setup for optical waveguide analysis.}
\centering
\label{fig:Writing-process}
\end{figure}

A depressed cladding waveguide is formed by scanning the sample multiple times with the fs-laser beam.
As illustrated in Fig. \ref{fig:Writing-process}(a), the cladding is built by offsetting the individual scans with respect to each other perpendicular to the writing direction to create a tubular arrangement of the tracks of reduced refractive index.
The center of the pattern, located \SI{200}{\micro\meter} beneath the sample surface, is left unexposed, leading to the formation of the waveguide core surrounded by the built cladding with a lower refractive index. 

For each set of experiments,
    multiple waveguides with different parameters were inscribed next to each other (Fig. \ref{fig:Writing-process}(b)).
Laser ablation and subsequent breaking were carried out to create flat end facets of the waveguides for light in- and out-coupling.
First, two laser ablation lines perpendicular to the waveguides were written on the surface of the sapphire sample using a high power setting of the laser and a repetition rate of \SI{60}{\kilo\hertz} (Fig. \ref{fig:Writing-process}(c)).
Each sample was broken along the ablation lines into a \SI{1}{\centi\meter}-long piece by applying force in the negative z-direction at one side of the ablation line (Fig. \ref{fig:Writing-process}(d)).
Subsequently, the broken sidewalls were prepared by grinding and polishing to achieve optical smoothness.
The obtained cross-sections provide direct access to the DCW end facets (Fig. \ref{fig:Writing-process}(c,d)).

\subsection{Optical waveguides analysis}
First, microscopic inspection of the waveguides was conducted from both above and via the edges of the sapphire substrates.
Afterwards, the mode profile as well as the losses of the waveguides were determined by optical characterization.
The mode profile gives information on the confinement properties of the waveguide, allowing to discriminate between single and multimode behavior.
Ideally, waveguides presenting circular mode profiles with near-Gaussian intensity distributions are preferred, as single mode operation is crucial for the vast majority of integrated optical applications, since it avoids inter-modal power transfers and allows a proper engineering of waveguide coupling. 

The measurement setup for optical waveguide analysis includes a light source that is coupled into the waveguide under test. 
The light is then guided through the waveguide and exits at the other end. 
The output light is detected using a microscope to observe the waveguide mode profile as well as with a powermeter to measure the transmitted power. 

\begin{figure}[t]
\includegraphics[width=1\textwidth]{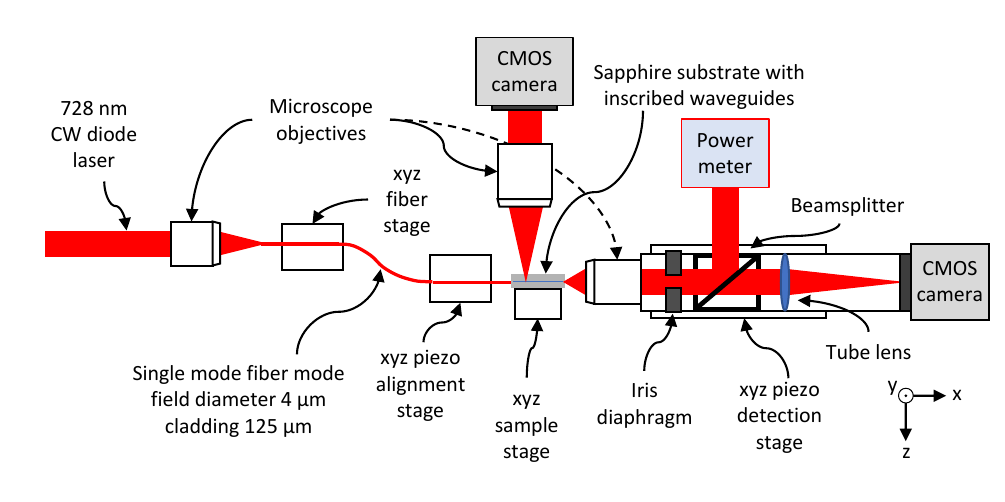}
\caption{Schematic representation of the waveguide characterization setup. Laser light from a \SI{728}{\nano\meter} diode laser is coupled into a single mode fiber via a microscope objective. This fiber is fixed on a xyz-Piezo stage for alignment to the sample.  A microscope images the fiber tip and incoupling site of the sample from above. The outgoing light from the sample is collected by a horizontal microscope mounted on a xyz-Piezo stage. It allows image capture via a CMOS camera as well as power measurement via a power meter.}
\centering
\label{fig:Characterization-setup}
\end{figure}

Fig. \ref{fig:Characterization-setup} shows a sketch of the measurement setup for optical waveguide analysis.
Laser light from a \SI{728}{\nano\meter} laser diode (Cobolt 06-MLD) is end-fire coupled into into a single-mode fiber (Thorlabs 630HP) using a microscope objective
    placed on an xyz-stage for manual adjustment.
The other end of the fiber is stripped and cleaved,
and the bare fiber tip is brought as close as possible
    to the waveguide end facet on the sidewall of the sapphire sample.
Consequently, the light from the bare single mode fiber is butt-coupled into the inscribed waveguides in the sapphire substrate
    placed on a xyz-stage for adjustment.
An additional video microscope offers a top view of the fiber-waveguide interface to aid the adjustment process.
The output light on the opposite end of the waveguide in the sapphire substrate is collected using a microscope objective with a numerical aperture of 0.5.
Stray light is minimized by an iris diaphragm and a beam-splitter divides the beam into two ports.
The first beam is measured by a powermeter to determine the waveguide losses.
The second beam is steered onto a CMOS camera
    to image the mode profile of the light propagating along the waveguide.
To optimize the positioning of fiber-butt-coupling as well as the collecting objective unit,
    including the power meter and the CMOS camera,
    three-axis ultra-high resolution piezo actuators (Newport NanoPZ), allowing a minimum incremental motion of \SI{30}{\nano\meter}, are used.
For determining the losses,
    the power of the light directly emitted from the fiber is measured as a reference
    and set into relation to the output power of the sapphire waveguide
        when coupled to the fiber.
This yields the total loss of the waveguide,
    which accounts for the coupling losses at input and output ports of the waveguide and for the propagation loss along the waveguide.

\section{Results and Discussion}

\subsection{Waveguide geometry optimization}
Circular depressed cladding structures were inscribed in sapphire 
    by positioning multiple laser tracks along a ring-shaped pattern.
Arranging the stages in a circular pattern leads to the formation of an elliptically shaped waveguide in sapphire
    due to the different refraction with varying writing depth.
Therefore,
    elliptically positioning patterns were investigated
        to compensate the ellipticity of the resulting waveguide.
Optimizing the waveguide geometry involved adjusting physical characteristics,
    including the dimension and shape of the waveguides, particularly the number of lines and the diameter of laser-inscribed ring-shaped pattern,
    as well as optimizing laser parameters: laser pulse energy, laser repetition rate or laser scanning speed.
The optimization process was repeated with single-mode profile and lowest total loss as targets.
Waveguides were written with fs-laser light at 524 nm with a polarization along the x-direction.
For focusing into the sapphire volume, 
    a microscope objective with a numerical aperture of 0.75 was used, 
        resulting in elliptical cross sections of single written lines with an estimated aspect ratio of 2.3.
The writing velocity was set to 15 mm/s.
The pulse duration, pulse energy and repetition rate were set to 300 fs, 65 or 78 nJ and 100 kHz, respectively.
The investigated waveguides were typically 1 cm long except for the cut-back measurements. 

\begin{figure}[t]
\includegraphics[width=0.9\textwidth]{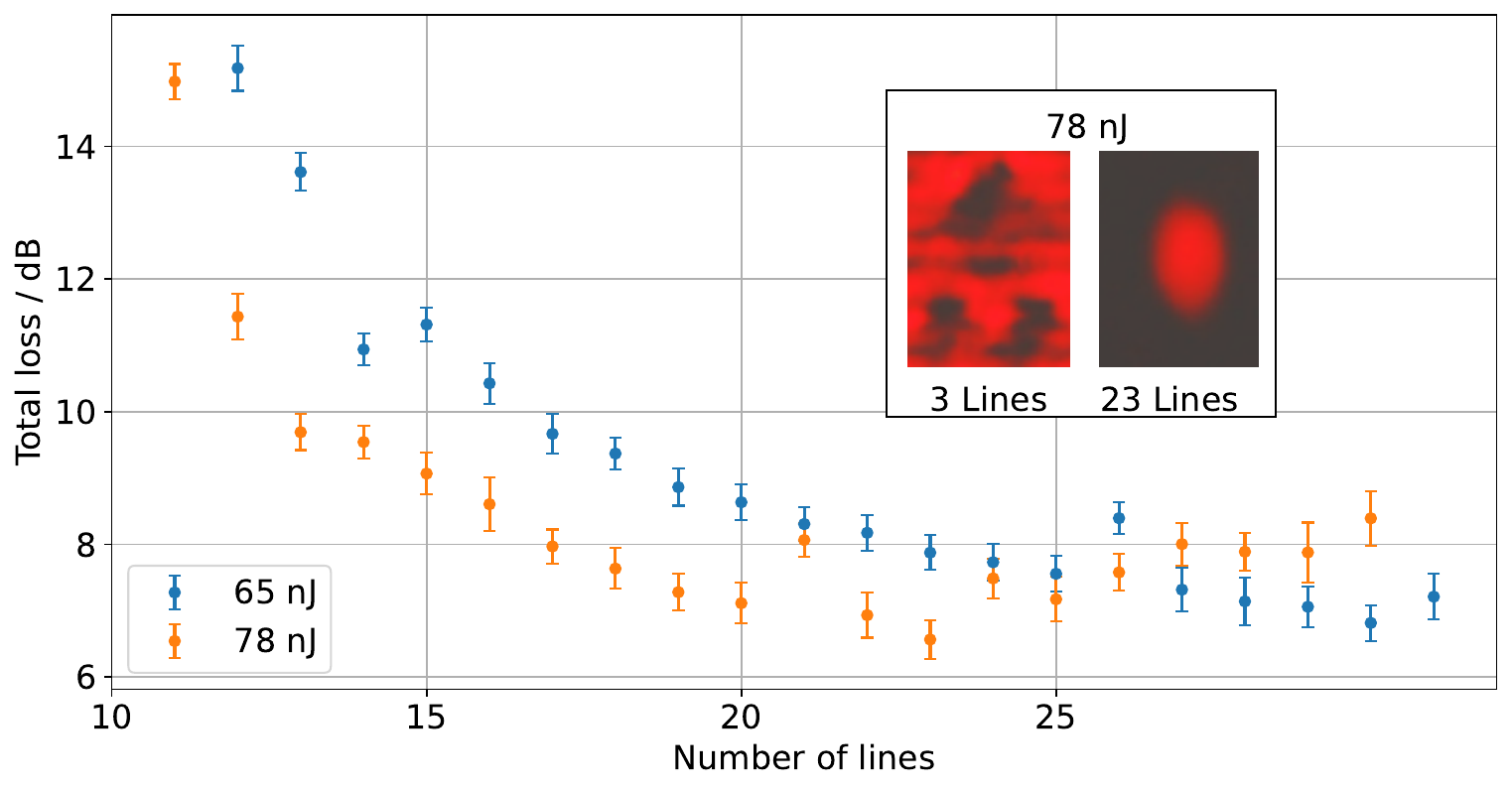}
\caption{Total loss as a function of the number of lines written forming the depressed cladding with laser pulse energies of 65 nJ (blue) and 78 nJ (orange).
Claddings with less than 10 lines did not show any wave-guiding properties.
Inset: Mode profile images for claddings written with $\SI{78}{nJ}$ and consisting of 3 lines and 23 lines.
Total loss was  \SI{15.2(3)}{dB} and \SI{6.5(2)}{dB} for 12 and 23 lines, respectively.}
\centering
\label{fig:Losses-lines}
\end{figure}

The depressed cladding geometry was varied in different aspects.
First, the number of lines that make up a ring-shaped cladding pattern with a radius of \SI{5}{\micro\meter} was varied from 3 to 30 lines.
Fig. \ref{fig:Losses-lines} shows the measured total loss depending on the number of lines written with laser pulse energies of \SI{65}{\nano\joule} (blue) and \SI{78}{\nano\joule} (orange).
Claddings consisting of less than 10 lines did not show any wave-guiding behavior.
A comparison of the profile of the waveguide with three lines in contrast to the Gaussian mode profile of the waveguide with 23 lines is shown in the inset of Fig. \ref{fig:Losses-lines}.
The measured total loss is highest for claddings consisting of 11 lines and decreases rapidly with increasing amount of lines, reaching a minimum of  \SI{6.6(2)}{dB} at 23 lines for a pulse energy of \SI{78}{\nano\joule} and \SI{6.8(2)}{dB} at 30 lines for a pulse energy of \SI{65}{\nano\joule}.
For a cladding consisting of 21 lines, the distance between lines is \SI{1.5}{\micro\meter} at the given radius of \SI{5}{\micro\meter}.
This distance between the lines was maintained for all further waveguides.

\setlength{\tabcolsep}{3pt} 
\renewcommand{\arraystretch}{1} 
\begin{center}
\begin{table}
\begin{tabular}{ c c c c c c c c}
 Radius/µm & 4 & 4.5 & 5 & 7.5 & 10 & 12.5 & 15 \\ \hline \hline
 65 nJ & \includegraphics[width=0.09\textwidth]{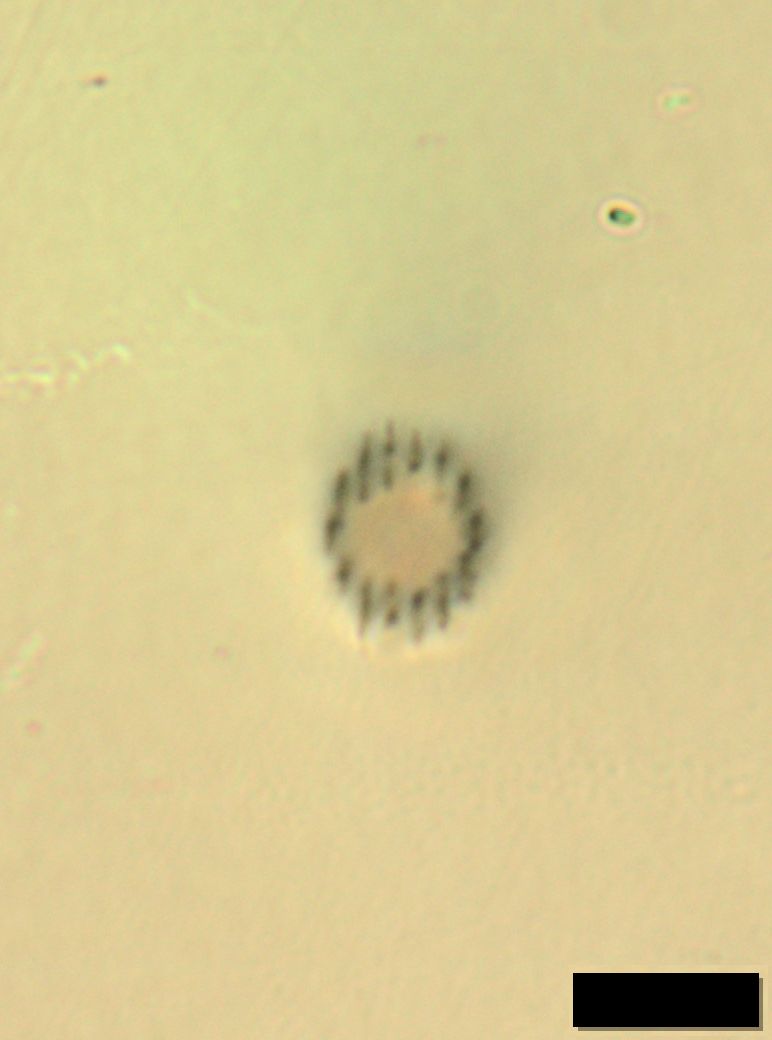} & \includegraphics[width=0.09\textwidth]{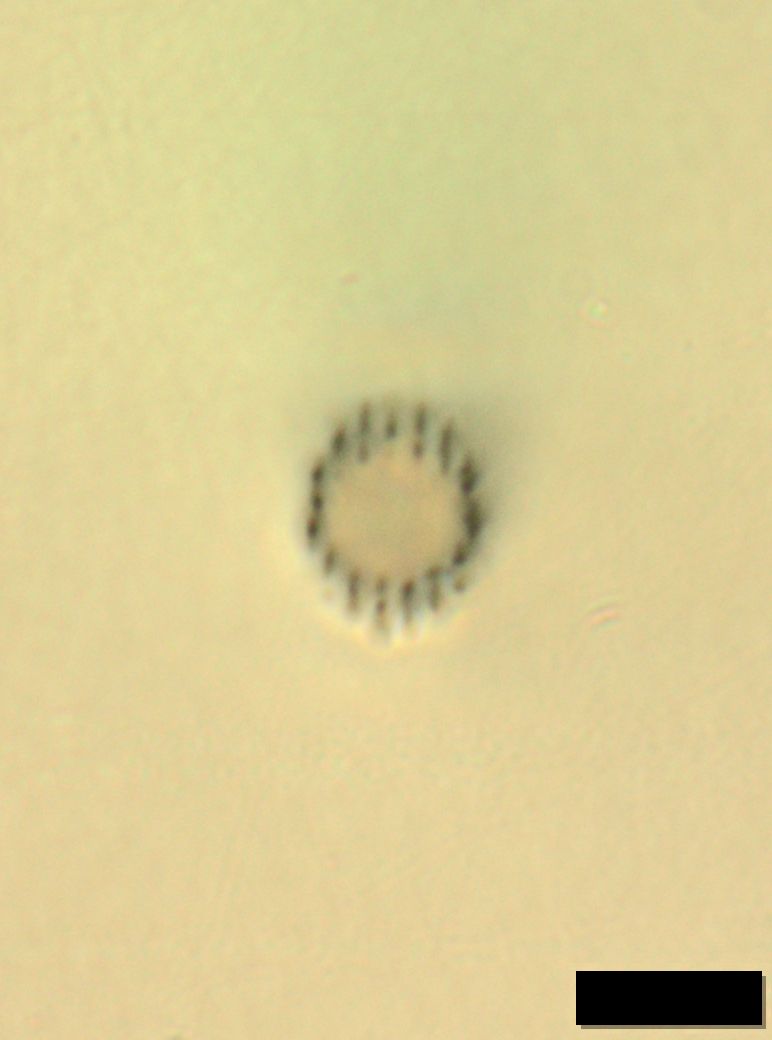} & \includegraphics[width=0.09\textwidth]{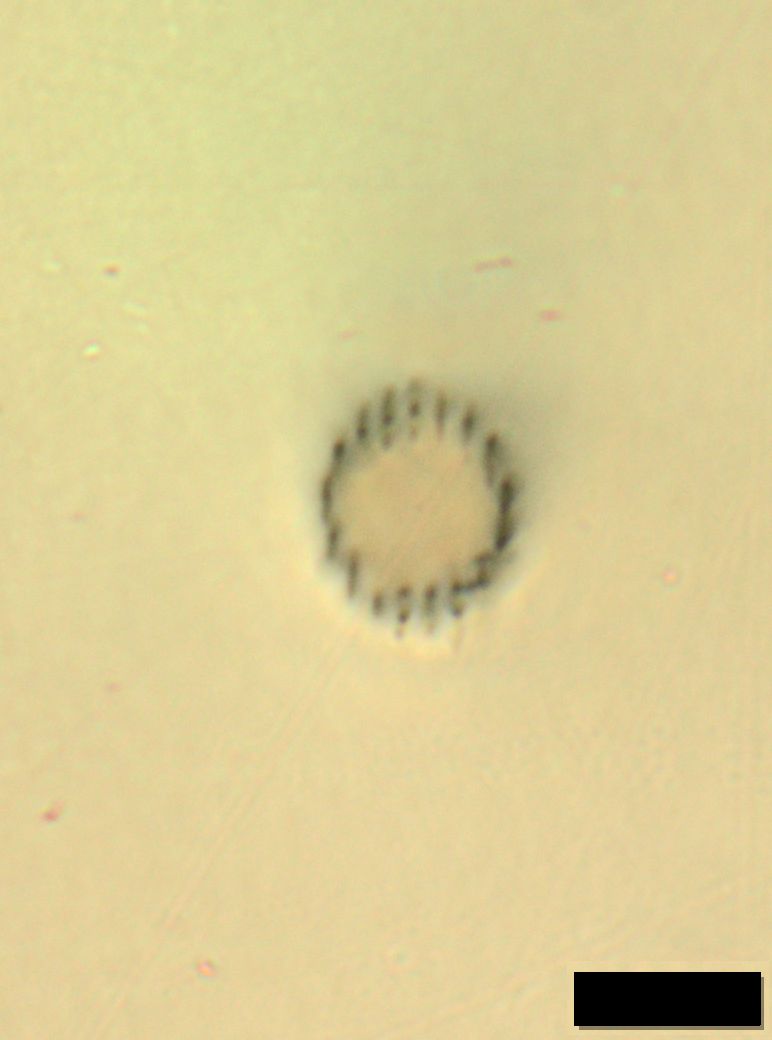} & \includegraphics[width=0.09\textwidth]{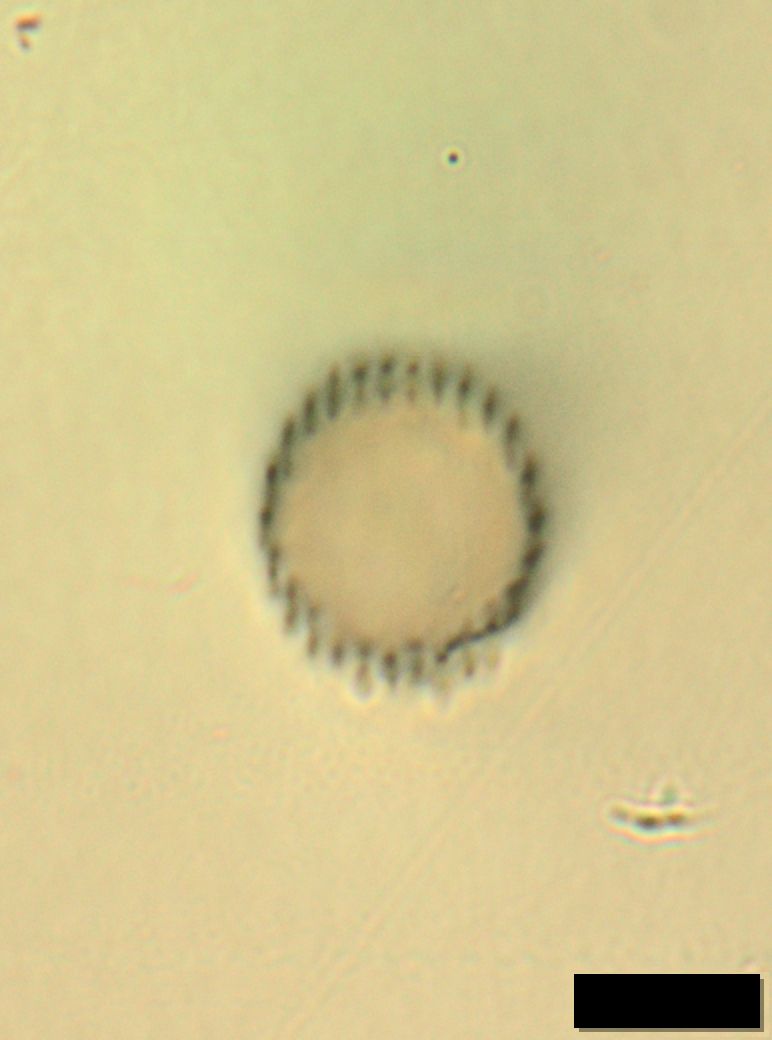} & \includegraphics[width=0.09\textwidth]{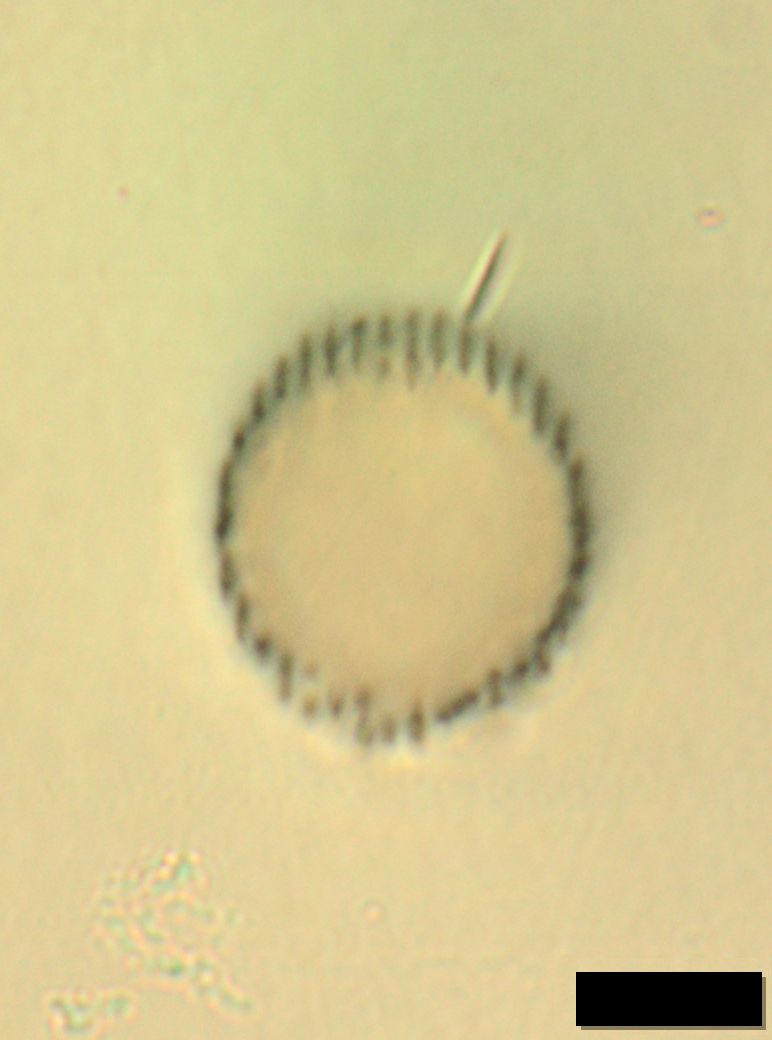} & \includegraphics[width=0.09\textwidth]{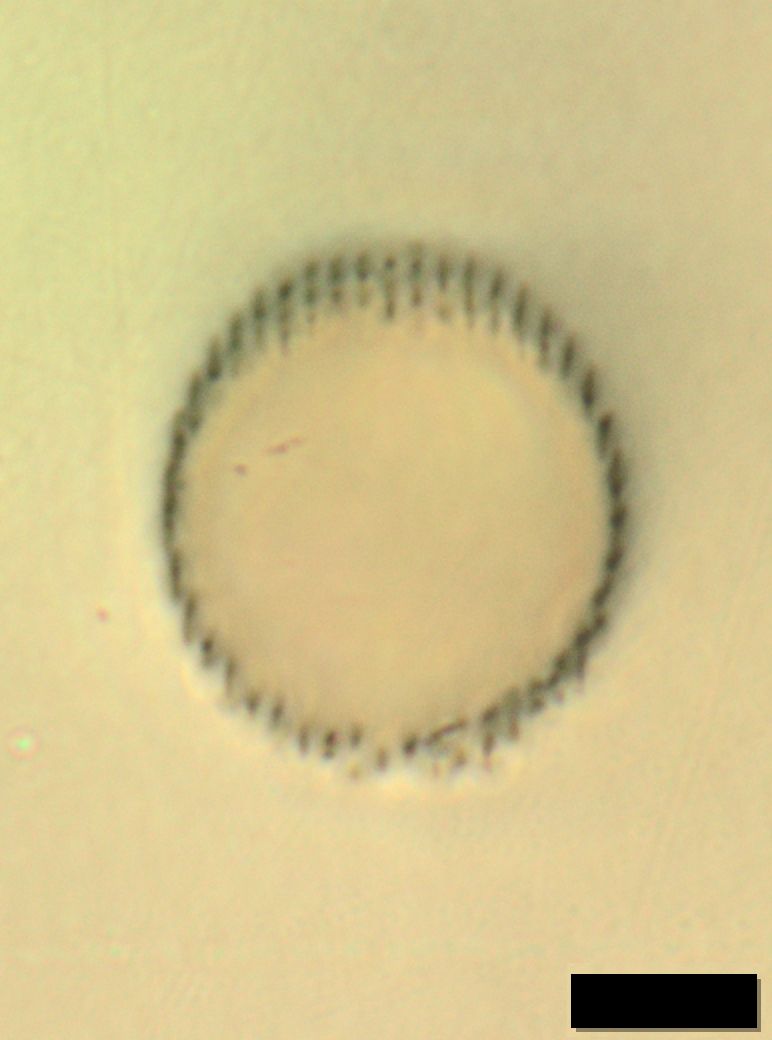} & \includegraphics[width=0.09\textwidth]{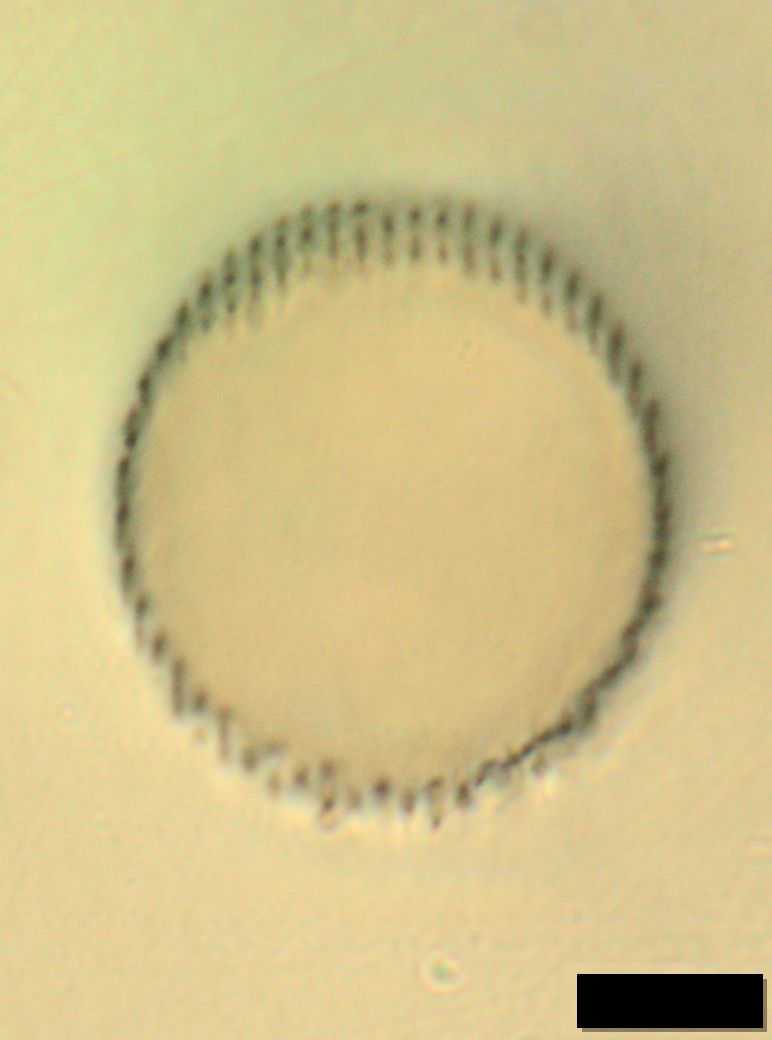} \\  
  & \includegraphics[width=0.09\textwidth]{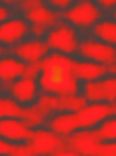} & \includegraphics[width=0.09\textwidth]{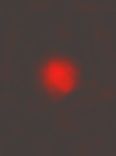} & \includegraphics[width=0.09\textwidth]{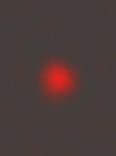} & \includegraphics[width=0.09\textwidth]{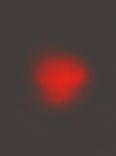} & \includegraphics[width=0.09\textwidth]{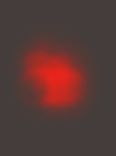} & \includegraphics[width=0.09\textwidth]{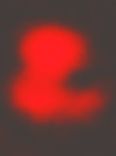} & \includegraphics[width=0.09\textwidth]{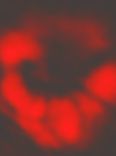} \\  
 Total loss/dB  & & 17.0 & 16.0  & 8.6 & 6.6 & 6.1 & 5.8 \\ \\ \hline
    78 nJ & \includegraphics[width=0.09\textwidth]{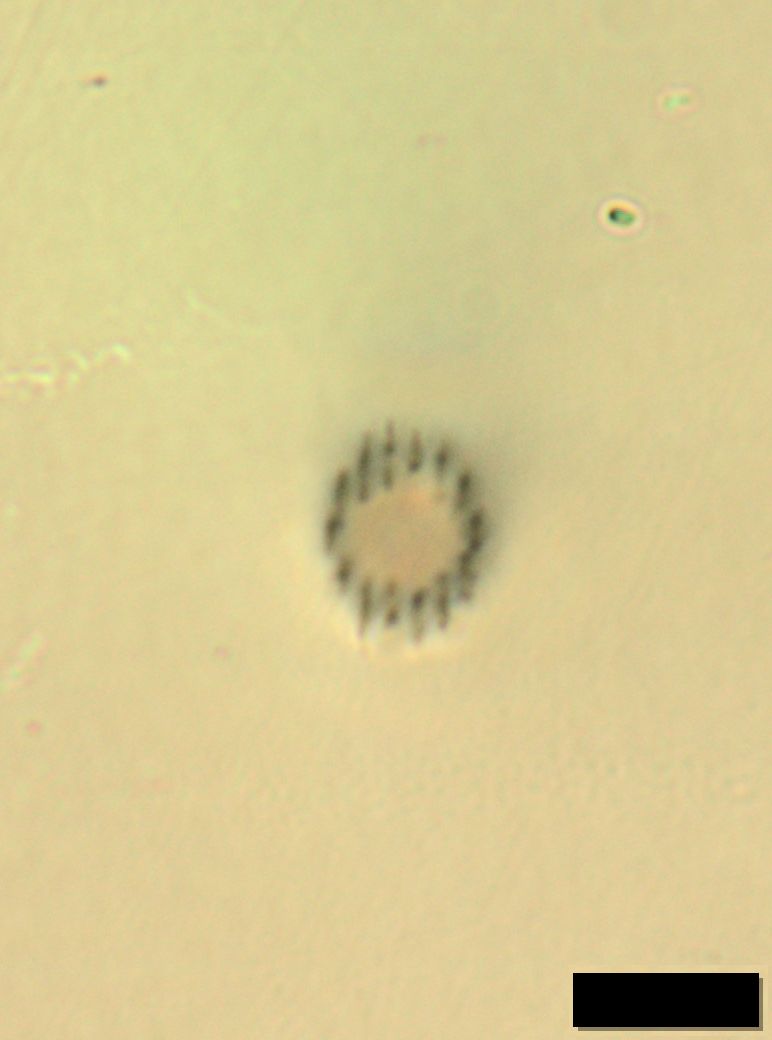} & \includegraphics[width=0.09\textwidth]{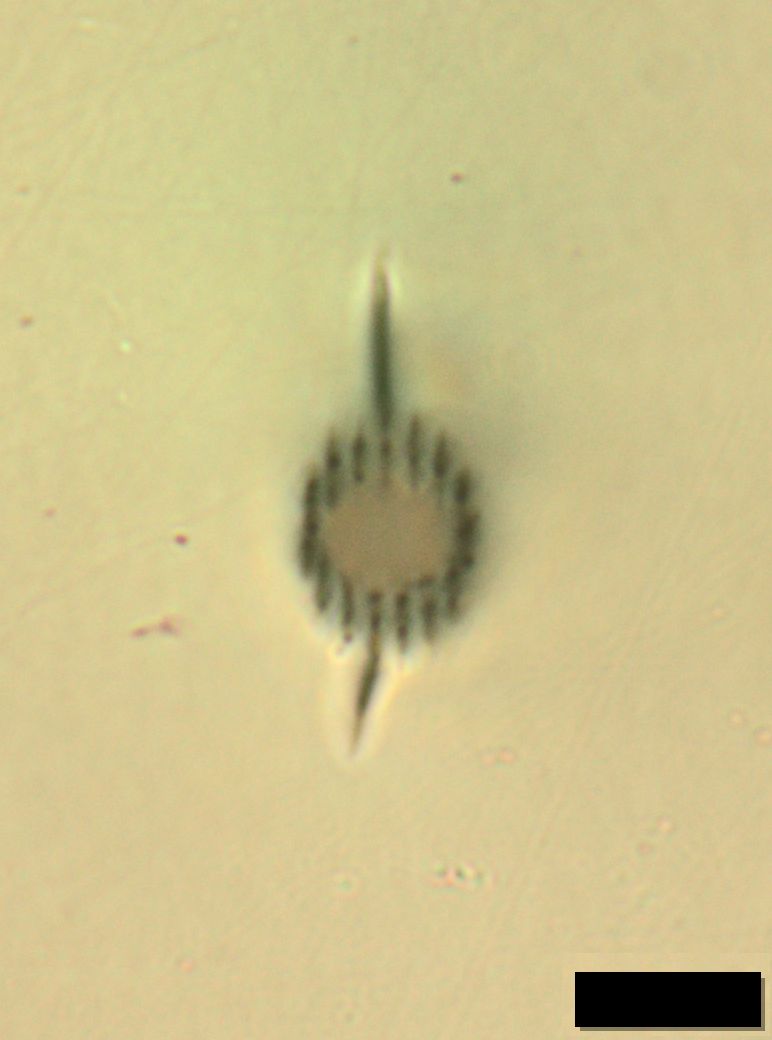} & \includegraphics[width=0.09\textwidth]{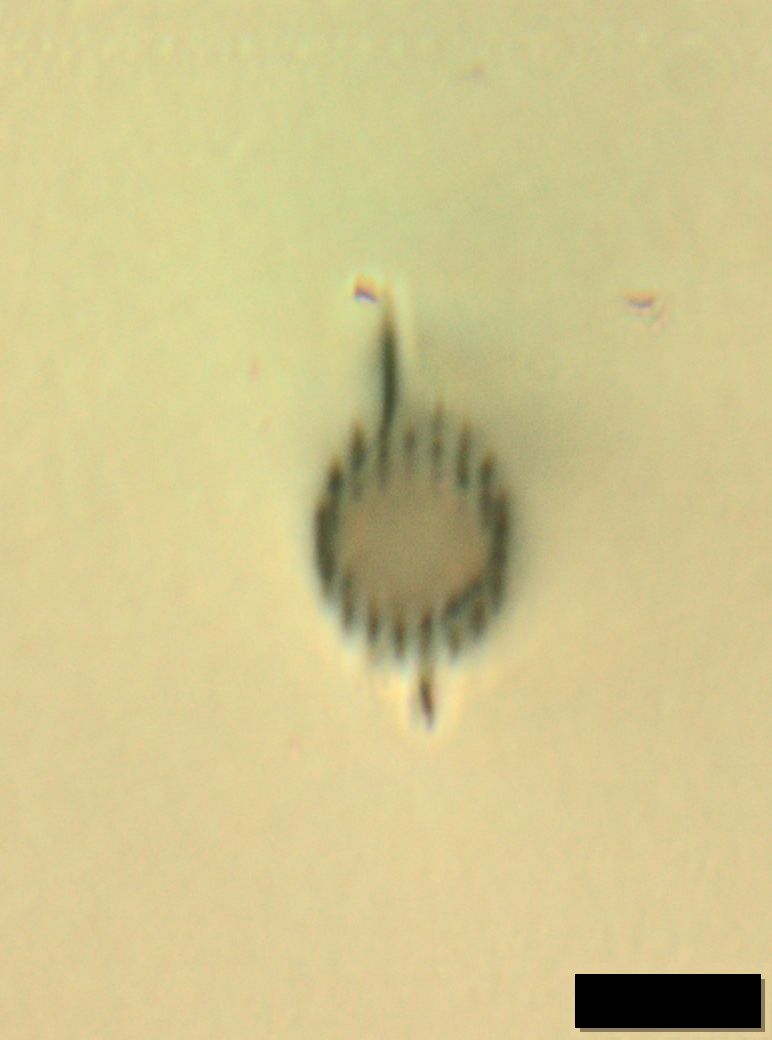} & \includegraphics[width=0.09\textwidth]{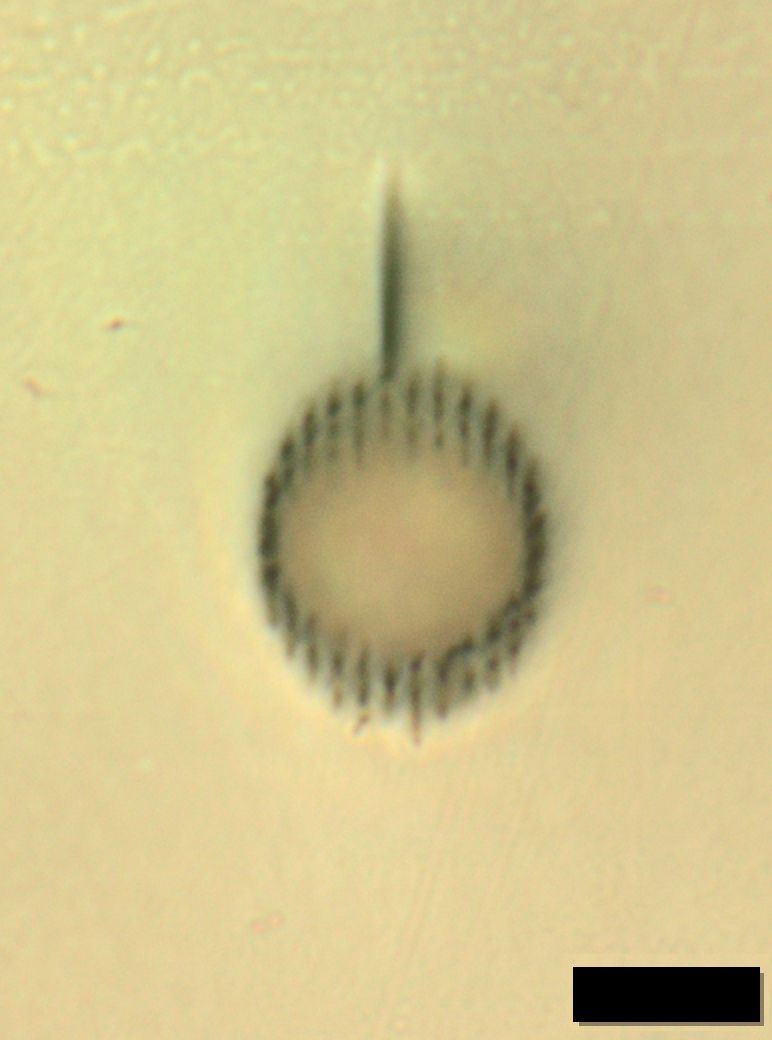} & \includegraphics[width=0.09\textwidth]{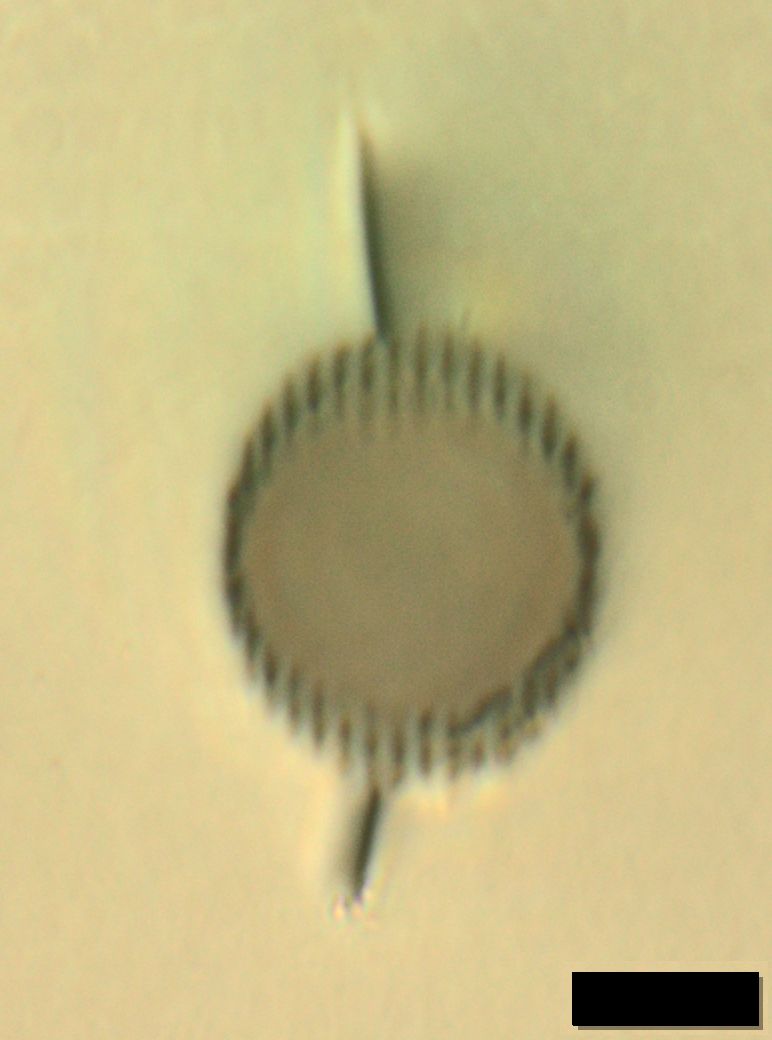} & \includegraphics[width=0.09\textwidth]{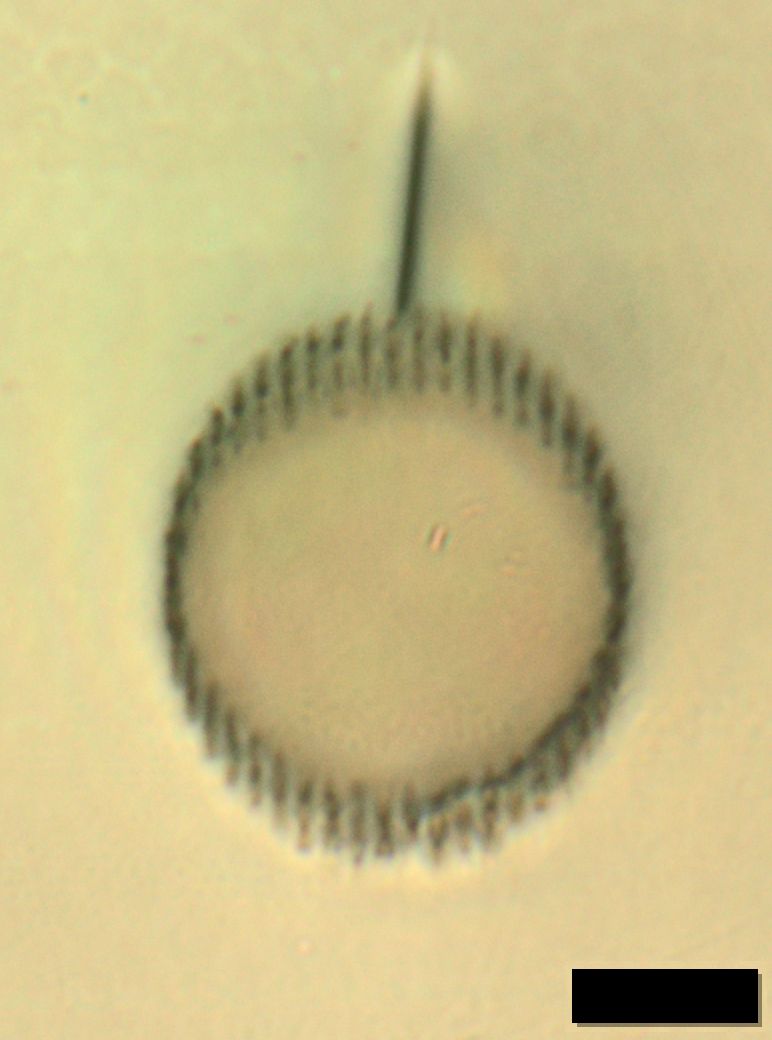} & \includegraphics[width=0.09\textwidth]{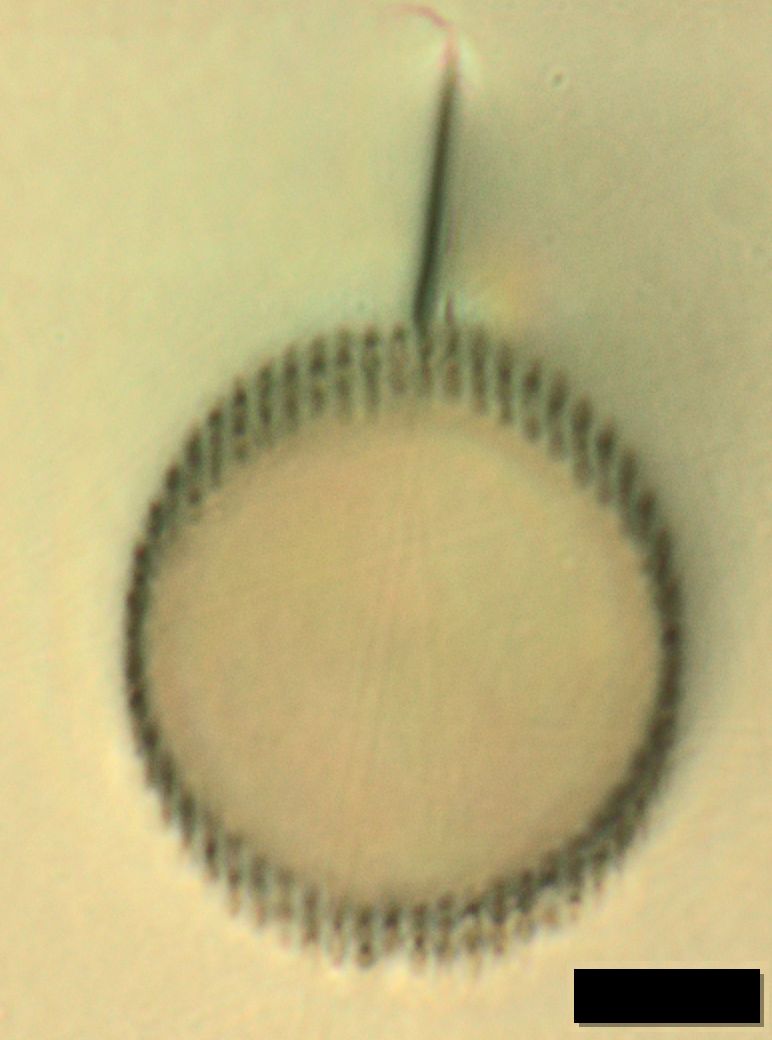} \\  
  & & \includegraphics[width=0.09\textwidth]{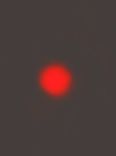} & \includegraphics[width=0.09\textwidth]{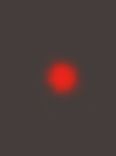} & \includegraphics[width=0.09\textwidth]{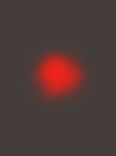} & \includegraphics[width=0.09\textwidth]{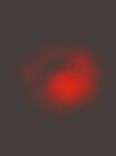} & \includegraphics[width=0.09\textwidth]{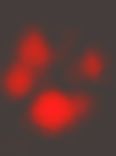} & \includegraphics[width=0.09\textwidth]{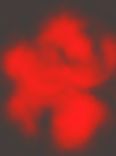} \\  
  Total loss/dB & & 14.4 & 12.9 & 8.3 & 6.4 & 5.6 & 5.1 \\
\end{tabular}
\caption{Microscope images (top, brown backdrop) and near-field images (bottom, black backdrop) of DCWs written with varying radii (4 - \SI{15}{\micro\meter}) at pulse energies of \SI{65}{\nano\joule} and \SI{78}{\nano\joule}. The black scale bar depicts \SI{10}{\micro\meter}.
The numbers at the bottom of each figure indicate the total loss for a \SI{10}{\milli\meter} long waveguide.}
\label{tab:WG-images}
\end{table}
\end{center}

Table \ref{tab:WG-images} displays waveguides with different core radii of 4 – \SI{15}{\micro\meter} fabricated again with pulse energies of 65 and 78 nJ.
On the top,
    exemplary microscope images of the DCW ends are shown,
        while on bottom the corresponding mode profiles are shown and total loss are indicated.
For waveguides with lines in a radius of \SI{4}{\micro\meter},
    no light guiding is observed.
For radii from \SI{4.5}{\micro\meter} to \SI{7.5}{\micro\meter},
    mode profiles reveal a near Gaussian shape robust against slight misalignments,
        indicating single mode wave-guiding.
For a core radius of \SI{10}{\micro\meter} and above,
    the images adopt more complex structures changing with the alignment of the end-fire coupled single mode fiber, 
        indicating multi-mode behavior.
Hereafter, the core radius is kept constant at \SI{7.5}{\micro\meter},
    showing the lowest total loss of \SI{8.6 \pm 0.2}{dB} and \SI{8.3 \pm0.2}{dB} at a pulse energy of \SI{65}{\nano\joule} and \SI{78}{\nano\joule}, respectively,
    while maintaining single-mode guiding.

\subsection{Curved depressed cladding waveguides}
To investigate the influence of the radius of curvature on the waveguide propagation characteristics, s-shaped curved waveguides with variable radii of curvature were fabricated and measured.
S-bend curves with curvature radii ranging from \SI{5}{\milli\meter} to \SI{60}{\milli\meter} were inscribed across the sample length of \SI{1}{\centi\meter}, using a pulse energy of \SI{65}{\nano\joule}.
Note that with a lower radius of curvature,
    the horizontal offset in y-direction between input and output and the total length of the waveguide are increased.

\begin{figure}[t]
\includegraphics[width=0.9\textwidth]{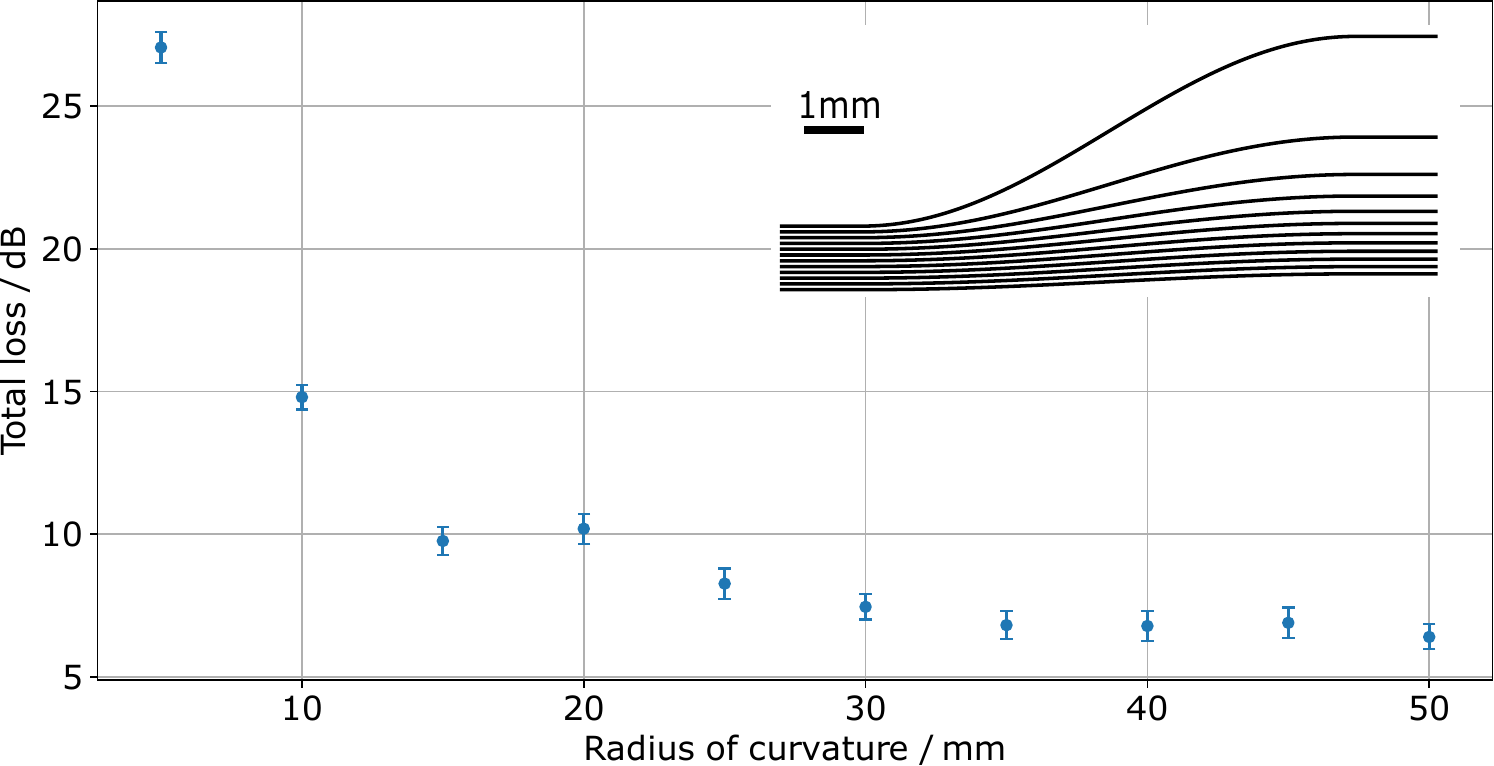}
\caption{Total loss of waveguides depending on their radius of curvature.
Inset: Top view of curves as drawn in the CAD program.
Curvature radius of \SI{5}{\milli\meter} (highest curvature, top) to \SI{60}{\milli\meter} (lowest curvature, bottom). 
Waveguides were
written with an ellipticity factor of 0.6 in the CAD program resulting in a circular pattern in Sapphire at a pulse energy of \SI{65}{\nano\joule}.
}
\centering
\label{fig:Curved-WG}
\end{figure}

Fig. \ref{fig:Curved-WG} shows the total loss as function of the radius of curvature of the waveguide.
With radii of curvature above \SI{35}{\milli\meter} ,
    the total loss falls within the range represented by two standard deviations of measurements of straight waveguides.
The total loss starts to increase with radii of curvature increasing from \SI{15}{\milli\meter} to \SI{35}{\milli\meter}  
    and a rapid increase is observed for radii smaller than \SI{15}{\milli\meter}.

\subsection{Propagation loss}
All loss values reported so far concern the total loss.
The total loss can be broken down into three components: in-coupling loss, propagation loss, and out-coupling loss.
In-coupling loss refers to the amount of light that is lost as it is coupled into the waveguide.
This loss can occur due to Fresnel reflection at the sample surface or due to an imperfect match between the waveguide mode and the light source´s mode.
Propagation loss refers to the amount of light that is lost as it propagates through the waveguide. This loss can occur due to absorption or scattering induced at material defects.
Out-coupling loss refers to the amount of light that is lost as it exits the waveguide and is largely determined by the Fresnel loss, as the characterization setup aims to collect the entire light field after the waveguide.

In general, the optimization of the waveguide geometry is done to minimize these losses.
The in-coupling and out-coupling losses can be minimized by accurately aligning the light source and detector with the waveguide, as well as using mode-matching techniques.
The propagation loss can be reduced by using a low-loss waveguide material, designing the waveguide to have a small mode area, and optimizing fs-laser writing parameters for writing tracks with low absorption and scattering cross sections.

To determine the contribution of the propagation loss, five identical straight waveguides (ring-shaped cladding pattern with a radius of \SI{5}{\micro\meter}) were written in a sapphire sample with a length of \SI{100}{\milli\meter}.
The length of the sample was gradually decreased by scribing, breaking and subsequent grinding and polishing.
The length error, \SI{\pm 0.1}{\milli\meter}, results from the uncertainty in grinding and polishing depth, after the breaking process.
After each breaking step, the total loss was measured for each of the five waveguides to perform a so-called cutback measurement. 

\begin{figure}[t]
\includegraphics[width=0.9\textwidth]{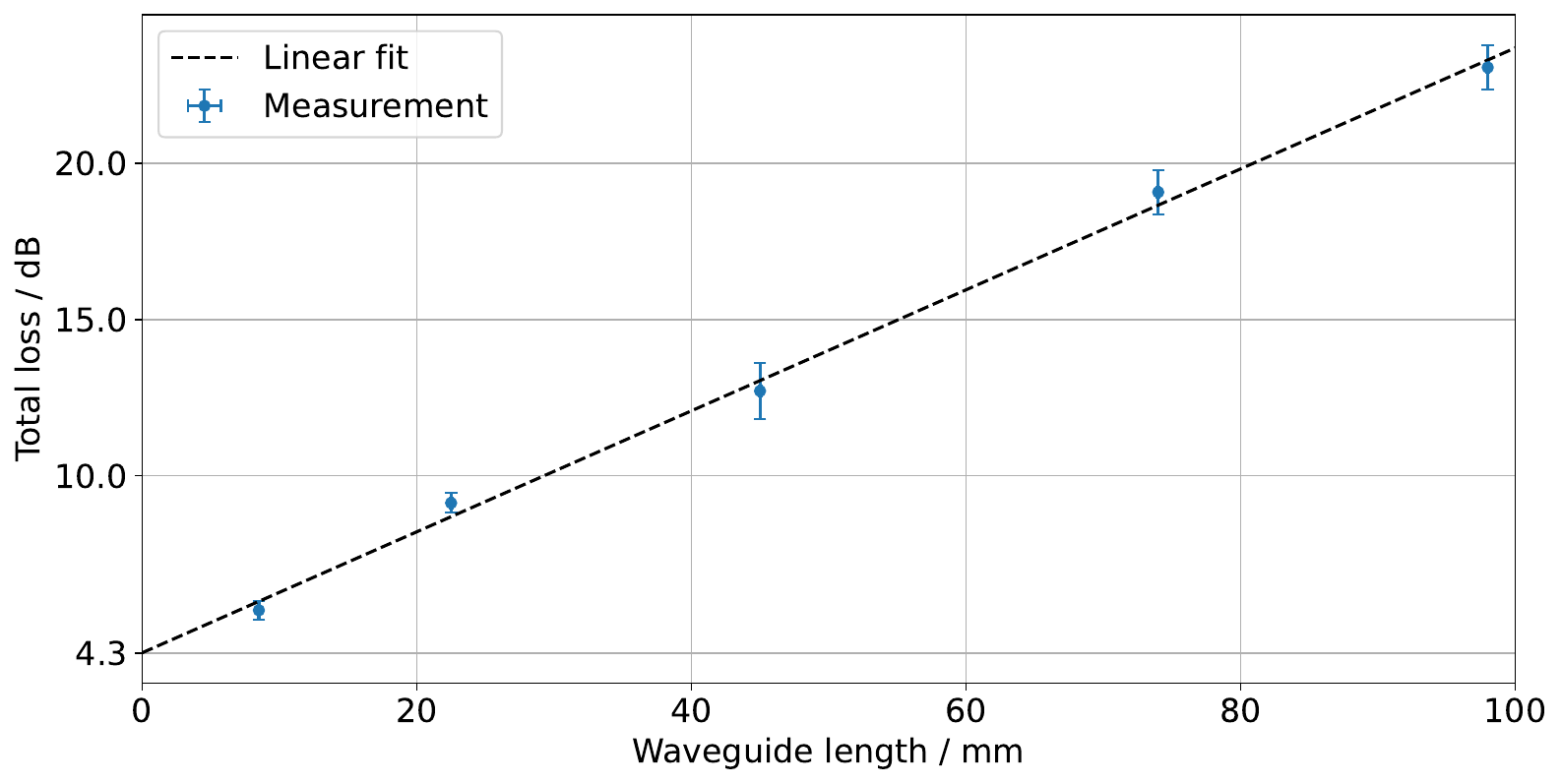}
\caption{Measured total loss in dependence of the waveguide length. The error of the waveguide length is \SI{\pm 0.1}{mm} and is smaller than the marker size.
The linear fit (dashed line) gives waveguide propagation loss of \SI{1.9 \pm 0.3}{dB\per cm} and in- and out-coupling losses of \SI{4.3 \pm 0.3}{dB}.}
\centering
\label{fig:Losses-length}
\end{figure}

Fig. \ref{fig:Losses-length} shows the measured total loss in dependence of the sample length.
The offset of the linear fit on the y-axis corresponds to the constant in- and out-coupling losses of \SI{4.3 \pm 0.3}{dB}, while its slope equals the propagation loss, which amounts to \SI{1.9 \pm 0.3}{dB \per \centi\meter} for the presented waveguides at the applied wavelength of \SI{728}{\nano\meter}.

\section{Conclusion and Outlook}
To the best of our knowledge, we demonstrated the first measurements with fs-laser written depressed cladding waveguides in sapphire with visible light, and of curved waveguides in undoped sapphire.
Utilizing a fs-laser enables precise control over the waveguide geometry and the ability
    to create highly localized modifications of refractive index in the bulk material,
        resulting in highly customizable waveguides through  adjustments to various structural and laser parameters.
We achieved transmission losses of \SI{1.9 \pm 0.3}{dB \per \centi\meter} at $\lambda = \SI{728}{\nano\meter}$ and coupling losses of \SI{4.3 \pm 0.3}{dB}.
These values enable the implementation of single-qubit and multi-qubit quantum gates with Calcium ions
    as shown with planar-fabricated integrated optics with similar values in \cite{Mehta2020}.
Advancements in laser technology and optimization techniques are anticipated to further enhance the quality of 
    the presented waveguides. 
    Furthermore, depressed cladding waveguides have applications for ionization, cooling, and repumping of trapped ions.
    The use of sapphire as a waveguide material offers several advantages
    including its high thermal conductivity, mechanical strength and chemical stability,
        which are important for many applications including trapped ion quantum computing and sensing.

\section{Funding}
Austrian Research Promotion Agency (FFG), project “OptoQuant” (37798980)

\section{Disclosures}
The authors declare no conflicts of interest.

\section{Data availability}
Data underlying the results presented in this paper are not publicly available at this time but may be obtained from the authors upon reasonable request.

\bibliographystyle{unsrt}  
\bibliography{refs.bib}

\begin{thebibliography}{10}

\bibitem{Haeffner2008}
H.~Häffner, C.~Roos, and R.~Blatt.
\newblock Quantum computing with trapped ions.
\newblock {\em Physics Reports}, 469(4):155--203, dec 2008.

\bibitem{Bruzewicz2019}
Colin~D. Bruzewicz, John Chiaverini, Robert McConnell, and Jeremy~M. Sage.
\newblock Trapped-ion quantum computing: Progress and challenges.
\newblock {\em Applied Physics Reviews}, 6(2):021314, jun 2019.

\bibitem{Blatt2012}
R.~Blatt and C.~F. Roos.
\newblock Quantum simulations with trapped ions.
\newblock {\em Nature Physics}, 8(4):277--284, apr 2012.

\bibitem{Brewer2019}
M.~Brewer, J.-S. Chen, A.~M. Hankin, E.~R. Clements, C.~W. Chou, D.~J.
  Wineland, D.~B. Hume, and D.~R. Leibrandt.
\newblock Al+ quantum logic clock with a systematic uncertainty below 1e-18.
\newblock {\em Physical Review Letters}, 123(3):033201, jul 2019.

\bibitem{Mehta2016}
Karan~K. Mehta, Colin~D. Bruzewicz, Robert McConnell, Rajeev~J. Ram, Jeremy~M.
  Sage, and John Chiaverini.
\newblock Integrated optical addressing of an ion qubit.
\newblock {\em Nature Nanotechnology}, 11(12):1066--1070, aug 2016.

\bibitem{Day2021}
Matthew~L Day, Kaushal Choonee, Zachary Chaboyer, Simon Gross, Michael~J
  Withford, Alastair~G Sinclair, and Graham~D Marshall.
\newblock A micro-optical module for multi-wavelength addressing of trapped
  ions.
\newblock {\em Quantum Science and Technology}, 6(2):024007, feb 2021.

\bibitem{McGuinness2022}
Hayden McGuinness, Michael Gehl, Craig Hogle, William~J. Setzer, Nickolas Karl,
  Nicholas Jaber, Justin Schultz, Joonhyuk Kwon, Megan Ivory, Rex Kay, Danield
  Dominguez, Douglous Trotter, Matt Eichenfield, and Daniel~L. Stick.
\newblock Integrated photonics for trapped ion quantum information experiments
  at sandia national laboratories.
\newblock In Mario Agio, Igor Aharonovich, Cesare Soci, and Matthew~T. Sheldon,
  editors, {\em Quantum Nanophotonic Materials, Devices, and Systems 2022}.
  {SPIE}, oct 2022.

\bibitem{Niffenegger2020}
R.~J. Niffenegger, J.~Stuart, C.~Sorace-Agaskar, D.~Kharas, S.~Bramhavar, C.~D.
  Bruzewicz, W.~Loh, R.~T. Maxson, R.~McConnell, D.~Reens, G.~N. West, J.~M.
  Sage, and J.~Chiaverini.
\newblock Integrated multi-wavelength control of an ion qubit.
\newblock {\em Nature}, 586(7830):538--542, oct 2020.

\bibitem{Vasquez2023}
Alfredo~Ricci Vasquez, Carmelo Mordini, Chlo{\'{e}} Verni{\`{e}}re, Martin
  Stadler, Maciej Malinowski, Chi Zhang, Daniel Kienzler, Karan~K. Mehta, and
  Jonathan~P. Home.
\newblock Control of an atomic quadrupole transition in a phase-stable standing
  wave.
\newblock {\em Physical Review Letters}, 130(13):133201, mar 2023.

\bibitem{Miura1997}
K.~Miura, Jianrong Qiu, H.~Inouye, T.~Mitsuyu, and K.~Hirao.
\newblock Photowritten optical waveguides in various glasses with ultrashort
  pulse laser.
\newblock {\em Applied Physics Letters}, 71(23):3329--3331, dec 1997.

\bibitem{Osellame2012}
Roberto Osellame, Giulio Cerullo, and Roberta Ramponi, editors.
\newblock {\em Femtosecond Laser Micromachining}.
\newblock Springer Berlin Heidelberg, 2012.

\bibitem{Tan2021}
Dezhi Tan, Zhuo Wang, Beibei Xu, and Jianrong Qiu.
\newblock Photonic circuits written by femtosecond laser in glass: improved
  fabrication and recent progress in photonic devices.
\newblock {\em Advanced Photonics}, 3(02), mar 2021.

\bibitem{Sun2022}
Bangshan Sun, Fyodor Morozko, Patrick~S. Salter, Simon Moser, Zhikai Pong,
  Raj~B. Patel, Ian~A. Walmsley, Mohan Wang, Adir Hazan, Nicolas Barr{\'{e}},
  Alexander Jesacher, Julian Fells, Chao He, Aviad Katiyi, Zhen-Nan Tian, Alina
  Karabchevsky, and Martin~J. Booth.
\newblock On-chip beam rotators, adiabatic mode converters, and waveplates
  through low-loss waveguides with variable cross-sections.
\newblock {\em Light: Science {\&} Applications}, 11(1), jul 2022.

\bibitem{Corrielli2021}
Giacomo Corrielli, Andrea Crespi, and Roberto Osellame.
\newblock Femtosecond laser micromachining for integrated quantum photonics.
\newblock {\em Nanophotonics}, 10(15):3789--3812, oct 2021.

\bibitem{Dong2013}
Ming-Ming Dong, Cheng-Wei Wang, Zheng-Xiang Wu, Yang Zhang, Huai-Hai Pan, and
  Quan-Zhong Zhao.
\newblock Waveguides fabricated by femtosecond laser exploiting both depressed
  cladding and stress-induced guiding core.
\newblock {\em Optics Express}, 21(13):15522, jun 2013.

\bibitem{Ren2016}
Yingying Ren, Yang Jiao, Javier R.~V{\'{a}}zquez de~Aldana, and Feng Chen.
\newblock Ti:sapphire micro-structures by femtosecond laser inscription:
  Guiding and luminescence properties.
\newblock {\em Optical Materials}, 58:61--66, aug 2016.

\bibitem{Berube2018}
Jean-Philippe B{\'{e}}rub{\'{e}}, Jerome Lapointe, Albert Dupont, Martin
  Bernier, and R{\'{e}}al Vall{\'{e}}e.
\newblock Femtosecond laser inscription of depressed cladding single-mode
  mid-infrared waveguides in sapphire.
\newblock {\em Opt. Lett.}, 44(1):37, dec 2018.

\bibitem{Romero2019}
Carolina Romero, Javier~Garc{\'{\i}}a Ajates, Feng Chen, and Javier
  R.~V{\'{a}}zquez de~Aldana.
\newblock Fabrication of tapered circular depressed-cladding waveguides in
  nd:{YAG} crystal by femtosecond-laser direct inscription.
\newblock {\em Micromachines}, 11(1):10, dec 2019.

\bibitem{Li2022}
Lingqi Li, Weijin Kong, and Feng Chen.
\newblock Femtosecond laser-inscribed optical waveguides in dielectric
  crystals: a concise review and recent advances.
\newblock {\em Advanced Photonics}, 4(02), mar 2022.

\bibitem{Chen2013}
Feng Chen and J.~R.~V{\'{a}}zquez de~Aldana.
\newblock Optical waveguides in crystalline dielectric materials produced by
  femtosecond-laser micromachining.
\newblock {\em Laser {\&} Photonics Reviews}, 8(2):251--275, may 2013.

\bibitem{Hempel2014}
Cornelius Hempel.
\newblock {\em Digital quantum simulation, Schrödinger cat state spectroscopy
  and setting up a linear ion trap}.
\newblock PhD thesis, University of Innsbruck, 2014.

\bibitem{Hellwig2010}
M~Hellwig, A~Bautista-Salvador, K~Singer, G~Werth, and F~Schmidt-Kaler.
\newblock Fabrication of a planar micro penning trap and numerical
  investigations of versatile ion positioning protocols.
\newblock {\em New Journal of Physics}, 12(6):065019, June 2010.

\bibitem{Daniilidis2011}
N~Daniilidis, S~Narayanan, S~A Möller, R~Clark, T~E Lee, P~J Leek, A~Wallraff,
  St~Schulz, F~Schmidt-Kaler, and H~Häffner.
\newblock Fabrication and heating rate study of microscopic surface electrode
  ion traps.
\newblock {\em New Journal of Physics}, 13(1):013032, jan 2011.

\bibitem{Allcock2013}
D.~T.~C. Allcock, T.~P. Harty, C.~J. Ballance, B.~C. Keitch, N.~M. Linke, D.~N.
  Stacey, and D.~M. Lucas.
\newblock A microfabricated ion trap with integrated microwave circuitry.
\newblock {\em Applied Physics Letters}, 102(4):044103, jan 2013.

\bibitem{Kunert2013}
P.~J. Kunert, D.~Georgen, L.~Bogunia, M.~T. Baig, M.~A. Baggash, M.~Johanning,
  and Ch. Wunderlich.
\newblock A planar ion trap chip with integrated structures for an adjustable
  magnetic field gradient.
\newblock {\em Applied Physics B}, 114(1–2):27--36, December 2013.

\bibitem{Berube2019}
Jean-Philippe Bérubé, Jerome Lapointe, Albert Dupont, Martin Bernier, and
  Réal Vallée.
\newblock Femtosecond laser inscription of depressed cladding single-mode
  mid-infrared waveguides in sapphire.
\newblock {\em Optics Letters}, 44:37--40, 2019.

\bibitem{Bai2012}
Jing Bai, Guanghua Cheng, Xuewen Long, Yishan Wang, Wei Zhao, Guofu Chen,
  Razvan Stoian, and Rongqing Hui.
\newblock Polarization behavior of femtosecond laser written optical waveguides
  in ti:sapphire.
\newblock {\em Optics Express}, 20(14):15035, jun 2012.

\bibitem{Wang2022}
Mohan Wang, Patrick~S. Salter, Frank~P. Payne, Adrian Shipley, Stephen~M.
  Morris, Martin~J. Booth, and Julian A.~J. Fells.
\newblock Single-mode sapphire fiber bragg grating.
\newblock {\em Optics Express}, 30(9):15482, apr 2022.

\bibitem{Kefer2022}
Stefan Kefer, Gian-Luca Roth, Julian Zettl, Bernhard Schmauss, and Ralf
  Hellmann.
\newblock Sapphire photonic crystal waveguides with integrated bragg grating
  structure.
\newblock {\em Photonics}, 9(4):234, apr 2022.

\bibitem{Mehta2020}
Karan~K. Mehta, Chi Zhang, Maciej Malinowski, Thanh-Long Nguyen, Martin
  Stadler, and Jonathan~P. Home.
\newblock Integrated optical multi-ion quantum logic.
\newblock {\em Nature}, 586(7830):533--537, oct 2020.

\end{thebibliography}

\end{document}